\pdfoutput=1
\documentclass[runningheads]{llncs}
\usepackage{graphicx}
\usepackage{cite}  %
\usepackage{letltxmacro}  %
\usepackage{amssymb}%
\usepackage{amsmath}%
\usepackage{booktabs}
\usepackage{footnote}%
\begin{document}
\title{A review: Deep learning for medical image segmentation using multi-modality fusion}
\footnotetext[1]{This paper has been published in the journal Array.}
\titlerunning{Deep learning for medical image segmentation using multi-modality fusion}
\author{Tongxue Zhou\inst{1,2} \and
St\'ephane Canu\inst{2}
\and Su Ruan\inst{1}}
\authorrunning{Zhou, Canu, Ruan.}
\institute{Universit\'e de Rouen Normandie, LITIS - QuantIF, Rouen 76183, France\\
\and
INSA Rouen, LITIS - Apprentissage, Rouen 76800, France}
\maketitle  

\begin{abstract}
Multi-modality is widely used in medical imaging, because it can provide multi-information about a target (tumor, organ or tissue). Segmentation using multi-modality consists of fusing multi-information to improve the segmentation. Recently, deep learning-based approaches have presented the state-of-the-art performance in image classification, segmentation, object detection and tracking tasks. Due to their self-learning and generalization ability over large amounts of data, deep learning recently has also gained great interest in multi-modal medical image segmentation. In this paper, we give an overview of deep learning-based approaches for multi-modal medical image segmentation task. Firstly, we introduce the general principle of deep learning and multi-modal medical image segmentation. Secondly, we present different deep learning network architectures, then analyze their fusion strategies and compare their results. The earlier fusion is commonly used, since it's simple and it focuses on the subsequent segmentation network architecture. However, the later fusion gives more attention on fusion strategy to learn the complex relationship between different modalities. In general, compared to the earlier fusion, the later fusion can give more accurate result if the fusion method is effective enough. We also discuss some common problems in medical image segmentation. Finally, we summarize and provide some perspectives on the future research.
\end{abstract}



\section{Introduction}

Segmentation using multi-modality has been widely studied with the development of medical image acquisition systems. Different strategies for image fusion, such as probability theory \cite{dubois1992combination}\cite{lapuyade2017segmenting}, fuzzy concept \cite{das2013neuro}\cite{balasubramaniam2014image}, believe functions \cite{smets1990combination}\cite{lian2019joint}, and machine learning \cite{vazquez2011segmentation}\cite{srivastava2012multimodal}\cite{cai2007probabilistic}\cite{zhang2011kernel} have been developed with success. For the methods based on the probability theory and machine learning, different data modalities have different statistical properties  which makes it difficult to model them using shallow models. For the methods based on the fuzzy concept, the fuzzy measure quantifies the degree of membership relative to a decision for each source. The fusion of several sources is achieved by applying the fuzzy operators to the fuzzy sets. For the methods based on the belief function theory, each source is first modeled by an evidential mass, the DempsterShafer rule is then applied to fuse all sources. The main difficulty to use the belief function theory and the fuzzy set theory relates to the choice of the evidential mass, the fuzzy measure and the fuzzy conjunction function. However, a deep learning-based network can directly encode the mapping. Therefore, the deep learning-based method has a great potential to produce better fusion results than conventional methods. Since 2012, several deep convolutional neural network models have been proposed such as AlexNet \cite{krizhevsky2012imagenet}, ZFNet \cite{zeiler2014visualizing}, VGG \cite{simonyan2014very}, GoogleNet \cite{szegedy2015going}, Residual Net \cite{he2016identity}, DenseNet \cite{huang2017densely}, FCN \cite{long2015fully} and U-Net \cite{ronneberger2015u}. These models have not only provided state-of-the-art performance for image classification, segmentation, object detection and tracking tasks, but also provide a new point of view for image fusion. There are mainly four reasons contributing to their success: Firstly, the main reason behind the amazing success of deep learning over traditional machine learning models is the advancements in neural networks, it learns high-level features from data in an incremental manner, which eliminates the need of domain expertise and hard feature extraction. And it solves the problem in an end to end manner. Secondly, the apprearance  of GPU and GPU-computing libraries  make the model can be trained 10 to 30 times faster than on CPUs. And the open source software packages provide efficient GPU implementations. Thirdly, publicly available datasets such as ImageNet, can be used for training, which allow researchers to train and test new variants of deep learning models. Finally, several available efficient optimization techniques also contributes the final success of deep learning, such as dropout, batch normalization, Adam optimizer and others, ReLU activation function and its variants, with that, we can update the weights and obtain the optimal performance.

\begin{figure}[htbp]
\centering
\includegraphics[height=6cm]{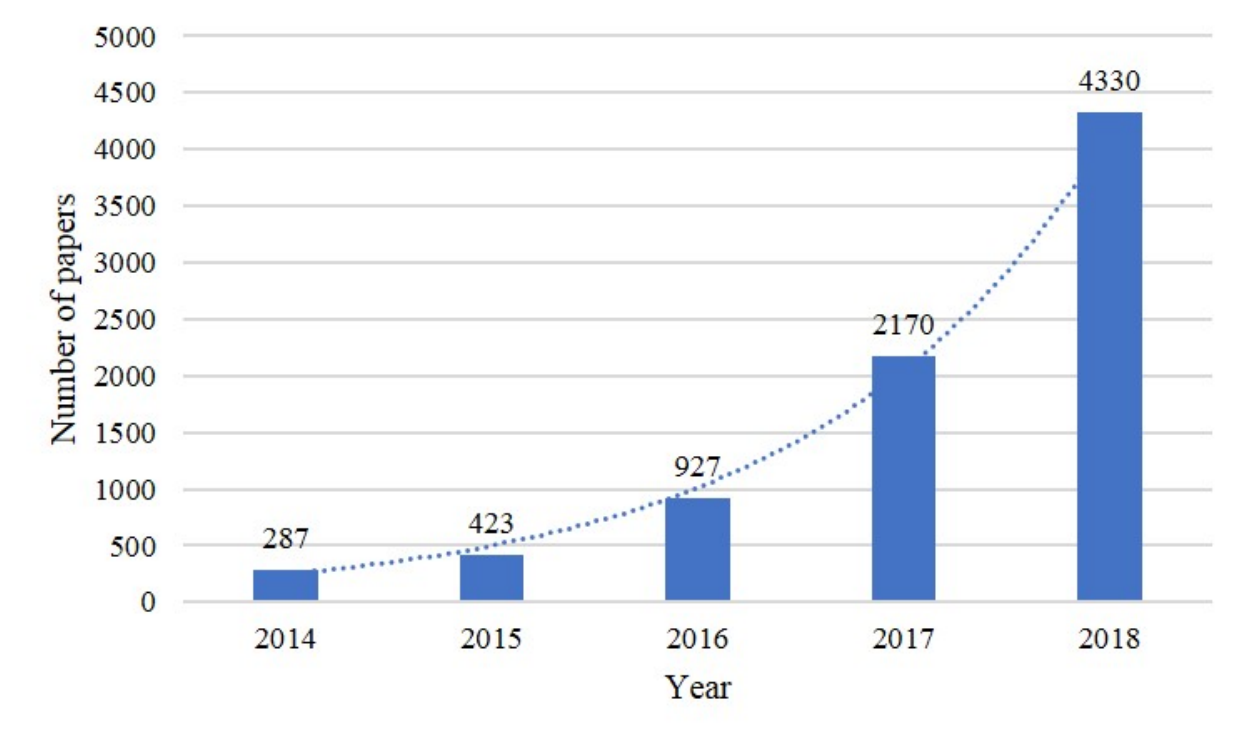}
\caption{The tendency of multi-modal medical image segmentation in deep learning}
\label{1}
\end{figure}

\begin{figure}[htbp]
\centering
\includegraphics[height=6cm,width=12cm]{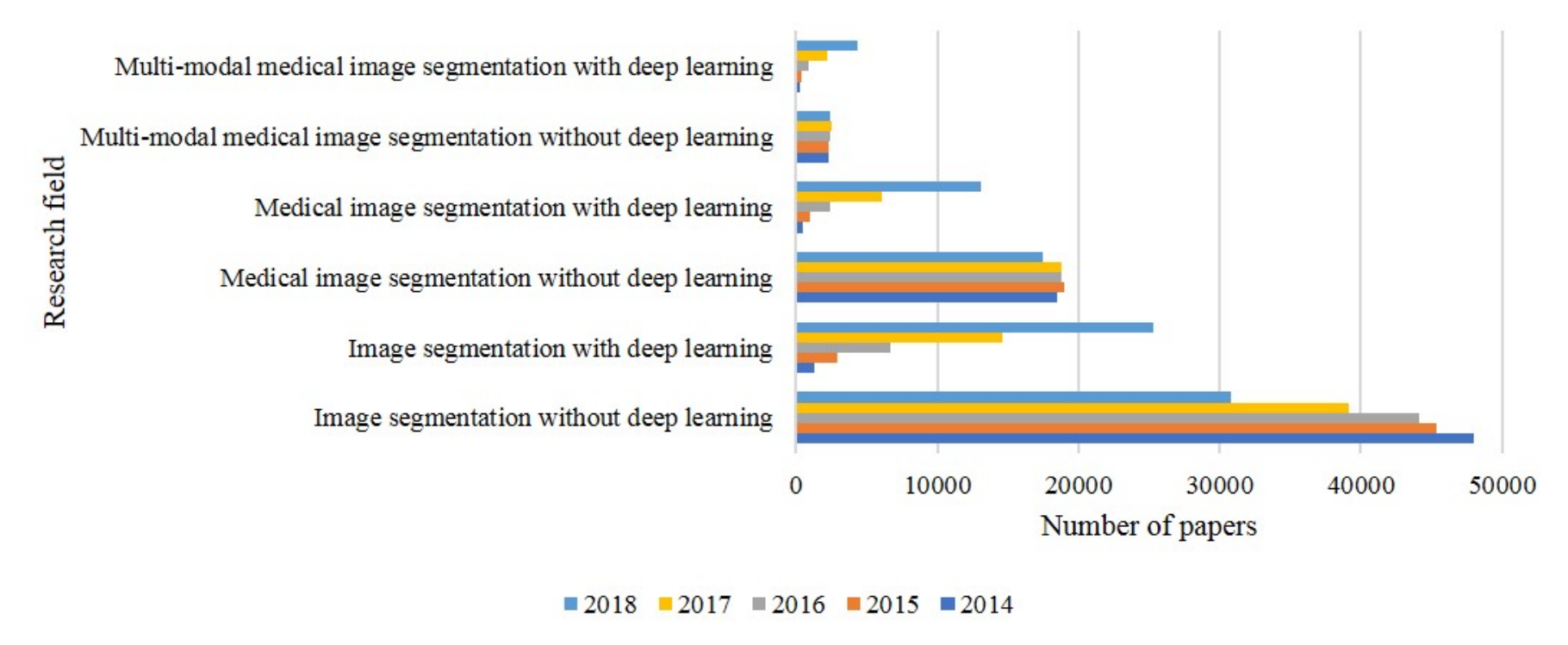}
\caption{The tendency of relative research field with/without deep learning}
\label{2}
\end{figure}

\begin{figure}[htbp]
\centering
\includegraphics[height=6cm]{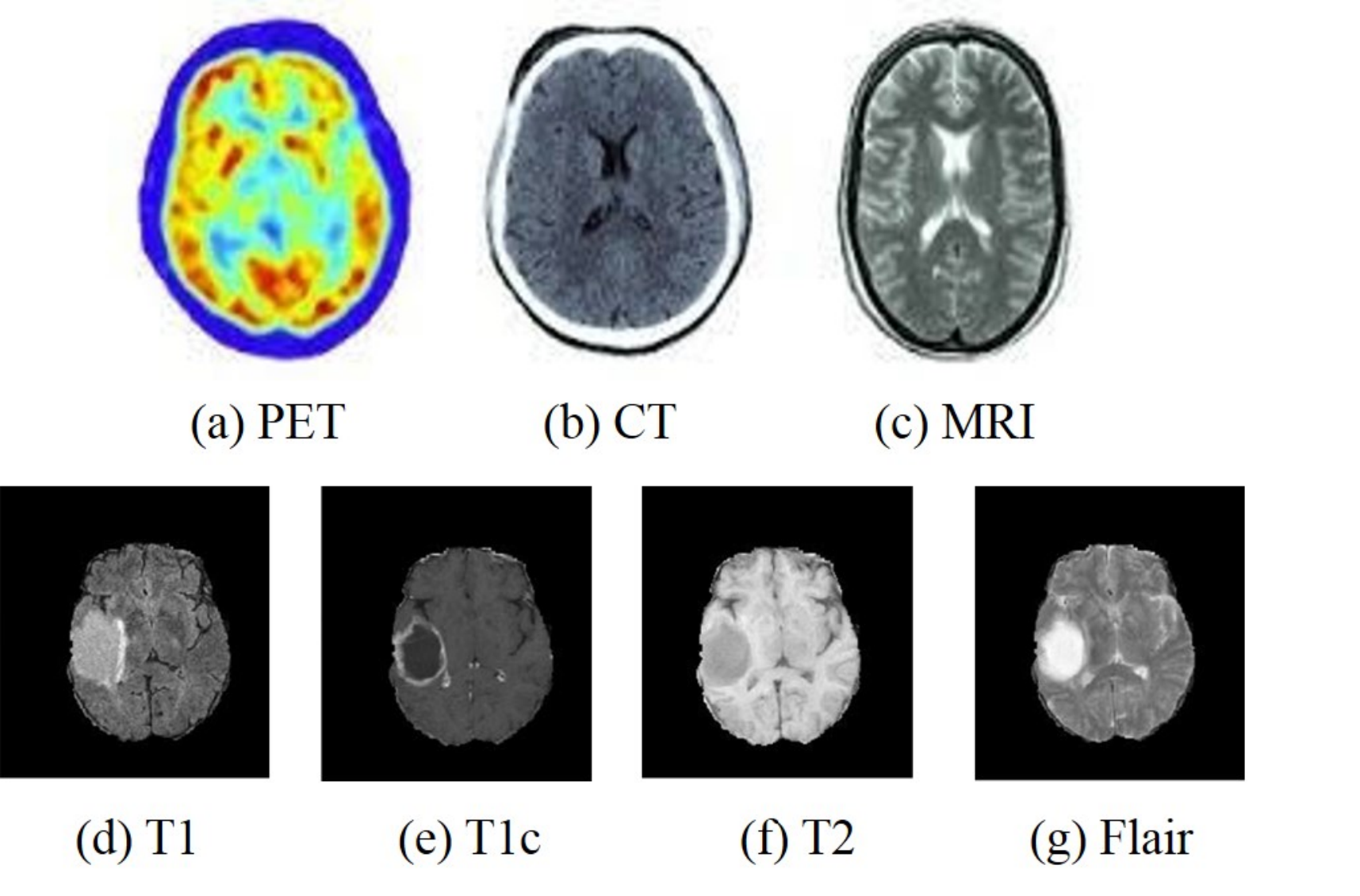}
\caption{The multi-modal medical images, (a)-(c) are the commonly used multi-modal medical images and (d)-(g) are the different sequences of brain MRI.}
\label{3}
\end{figure}

Motivated by the success of deep learning, researches in medical image field have also attempted to apply deep learning-based approaches to medical image segmentation in the brain \cite{havaei2017brain}\cite{pereira2016brain}\cite{menze2015multimodal}, lung \cite{kalinovsky2016lung}, pancreas \cite{fu2018hierarchical}\cite{roth2015deeporgan}, prostate \cite{yu2017volumetric} and multi-organ \cite{zhou2017deep}\cite{trullo2017joint}. Medical image segmentation is an important area in medical image analysis and is necessary for diagnosis, monitoring and treatment. The goal is to assign the label to each pixel in images, it generally includes two phases, firstly, detect the unhealthy tissue or areas of interest; secondly, decliner the different anatomical structures or areas of interest. These deep learning-based methods have achieved superior performance compared to traditional methods in medical image segmentation task. In order to obtain more accurate segmentation for better diagnosis, using multi-modal medical images has been a growing trend strategy. A thorough analysis of the literature with the keywords ‘deep learning', 'medical image segmentation' and 'multi-modality' on Google Scholar search engine is performed in Figure \ref{1}, which is queired on July 17, 2019. We can observe that the number of papers increases every year from 2014 to 2018, which means multi-modal medical image segmentation in deep learning are obtaining more and more attention in recent years. To have a better understanding of the dimension of this research field, we compare the scientific production of the image segmentation community, the medical image segmentation community, and the medical image segmentation using multi-modality fusion with and without deep learning in Figure \ref{2}. From the figure we can see, the amount of papers has a descent or even tendency in the methods without deep learning, but there is an increase number of papers using deep learning method in every research field. Especially in medial image segmentation field, due to the limited datasets, classical methods take still a more dominant position, but we can see an obvious increasing tendency in the methods using deep learning. The principal modalities in medical images analysis are computed tomography (CT), magnetic resonance imaging (MRI) and positron emission tomography (PET). Compared to single images, multi-modal images help to extract features from different views and bring complementary information, contributing to better data representation and discriminative power of the network. As pointed out in \cite{bhatnagar2015new}, the CT image can diagnose muscle and bone disorders, such as bone tumors and fractures, while the MR image can offer a good soft tissue contrast without radiation. Functional images, such as PET, lack anatomical characterization, while can provide quantitative metabolic and functional information about diseases. MRI modality can provide complementary information due to its dependence on variable acquisition parameters, such as T1-weighted (T1), contrast-enhanced T1-weighted (T1c), T2-weighted (T2) and Fluid attenuation inversion recovery (Flair) images. T2 and Flair are suitable to detect the tumor with peritumoral edema, while T1 and T1c to detect the tumor core without peritumoral edema. Therefore, applying multi-modal images can reduce the information uncertainty and improve clinical diagnosis and segmentation accuracy \cite{guo2019deep}. Several widely used multi-modal medical images are described in Figure \ref{3}. The earlier fusion is simple and most works use the fusion strategy to do the segmentation, it focuses on the subsequent complex segmentation network architecture designs, but it doesn't consider the relationship between different modalities and doesn't analyze how to fuse the different feature information to improve the segmentation performance. However, the later fusion pays more attention on the fusion problem, because each modality is employed as an input of one network which can learn complex and complementary feature information of each modality. In general, compared to the earlier fusion, the later fusion can achieve better segmentation performance if the fusion method is effective enough. And the selection of fusion method depends on the specific problem.

There are also some other reviews on medical image analysis using deep learning. However, they don't focus on the fusion strategy. For example, Litjens et al. \cite{litjens2017survey} reviewed the major deep learning concepts in medical image analysis. Bernal et al. \cite{bernal2018deep} gave an overview in deep CNN for brain MRI analysis. In this paper, we focus on fusion methods of multi-modal medical images for medical image segmentation.
  
The rest of the paper is structured as followed. In Section 2 we introduce the general principle of deep learning and multi-modal medical image segmentation. In Section 3, we present how to prepare the data before feeding to the network. In Section 4, we describe the detailed multi-modal segmentation network based on different fusion strategies. In Section 5, we discuss some common problems appeared in the field. Finally, we summarize and discuss the future perspective in the field of multi-modal medical image segmentation. 


\section{Deep learning based methods}
\subsection{Deep learning}
Deep learning refers to a neural network with multiple layers of nonlinear processing units \cite{lecun2015deep}. Each successive layer uses the output from the previous layer as input. The network can extract the complex hierarchy features from a large amount of data by using these layers. In recent years, deep learning has made significant improvements in image classification, recognition, object detection and medical image analysis, where they have produced excellent results comparable to or sometime superior to human experts. Among the known deep learning algorithms, such as stacked auto-encoders \cite{bengio2007greedy}, deep Boltzmann machines \cite{salakhutdinov2009deep}, and convolutional neural networks\cite{lecun1989backpropagation}, the most successful one for image segmentation is convolutional neural networks (CNN). It was first proposed in 1989 by LeCun  and the first successful real-world application \cite{lecun1998gradient} is the hand-written digit recognition in 1998 by LeCun, where he presented a five-layer fully-adaptive architecture. Due to its accuracy results (1\% error rate and 9\% reject rate from a dataset of 2007 handwritten characters), the neural networks can be applied into a real-world problem. However, it did not gather much attention until the contribution of Krizhevsky et al. to the ImageNet challenge in 2012. The proposed AlexNet \cite{krizhevsky2012imagenet}, similar to LeNet but deeper, outperformed all the competitors and won the challenge by reducing the top-5 error (the percentage of test examples for which the correct class was not in the top 5 predicted classes) from 26\% to 15.3\%. In the subsequent years, other based on CNN architectures are proposed, including VGGNet \cite{simonyan2014very}, GoogleNet \cite{szegedy2015going}, Residual Net \cite{he2016identity} and DenseNet \cite{huang2017densely}, Table \ref{Table.1} describes the details of these network architectures.

\begin{table}
\begin{center}
\resizebox{\textwidth}{!}{%
\begin{tabular}{ccccc}
\toprule 
\textbf{Architecture}& \textbf{Article}& \textbf{Rank on ILSVRC}& \textbf{Top-5 error rate}& \textbf{Number of parameters}\\ 
\midrule
LeNet\cite{lecun1998gradient}& LeCun et al. 1998 & N/A& N/A& 60 thousand\\ 
AlexNet\cite{krizhevsky2012imagenet}& Krizhevsky et al. 2012& 1st& 16.4\% & 60 million\\ 
ZFNet\cite{zeiler2014visualizing}& Zeiler et al. 2013& 1st& 11.7\% & N/A\\ 
VGG Net\cite{simonyan2014very}& Simonyan et al. 2014& 2nd& 7.3\% & 138 million\\ 
GoogleNet\cite{szegedy2015going}& Szegedy et al. 2015& 1st& 6.7\% & 5 million (V1) \& 23 million (V2)\\
ResNet\cite{he2016identity}& He. Kaiming et al. 2016& 1st& 3.57\% & 25.6 million (ResNet-50)\\ 
DenseNet\cite{huang2017densely}& Huang et al. 2017& N/A& N/A& 6.98 million (DenseNet-100, k=12)\\ 
\bottomrule
\end{tabular}
}
\end{center}
\caption{Summary of deep learning network architectures, ILSVRC:
ImageNet Large Scale Visual Recognition Challenge.}\label{Table.1}
\end{table}

CNN is a multi-layer neural network containing convolution, pooling, activation and fully connected layers. Convolution layers are the core of CNNs and are used for feature extraction. The convolution operation can produce different feature maps depending on the filters used. Pooling layer performs a downsampling operation by using maximum or average of the defined neighbourhood as the value to reduce the spatial size of each feature map. Non-linear rectified layer (ReLU) and its modifications such as Leaky ReLU are among the most commonly used activation functions \cite{he2015delving}, which transforms data by clipping any negative input values to zero while positive input values are passed as output. Neurons in a fully connected layer are fully connected to all activations in the previous layer. They are placed before the classification output of a CNN and are used to flatten the results before a prediction is made using linear classifiers. While training the CNN architecture, the model predicts the class scores for training images, computes the loss using the selected loss function and finally updates the weights using the gradient descent method by back-propagation. The cross-entropy loss is one of the most widely used loss functions and stochastic gradient descent (SGD) is the most popular method to operate gradient descent. 

\subsection{Multi-modal medical image segmentation}
Due to the variable size, shape and location of target tissue, medical image segmentation is one of the most challenging tasks in the field of medical image analysis. Despite the variety of proposed segmentation network architectures, it is still hard to compare the performance of different algorithms, because most of the algorithms are evaluated on different sets of data and reported in different metrics. In order to obtain accurate segmentation and compare different state-of-the-art methods, some well-known publicly challenges for segmentation are created, such as Brain Tumour Segmentation (BraTS) \cite{menze2015multimodal}, Ischemic Stroke Lesion Segmentation(ISLES)\footnote{http://www.isles-challenge.org}, MR Brain Image Segmentation (MRBrainS) \cite{mendrik2015mrbrains}, Neonatal Brain Segmentation (NeoBrainS) \cite{ivsgum2015evaluation}, Combined (CT-MR) Healthy Abdominal Organ Segmentation (CHAOS)\footnote{https://chaos.grand-challenge.org}, 6-month infant brain MRI Segmentation (Iseg-2017) \cite{wang2019benchmark} and  Automatic intervertebral disc localization and segmentation from 3D Multi-modality MR (M3) Images (IVDM3Seg)\footnote{https://ivdm3seg.weebly.com}. Table \ref{Table.2} describes the detailed dataset information mentioned above. Table \ref{Table.3} shows the main evaluation metrics in these datasets.

\begin{figure}[htbp]
\centering
\includegraphics[width=12cm, height=2cm]{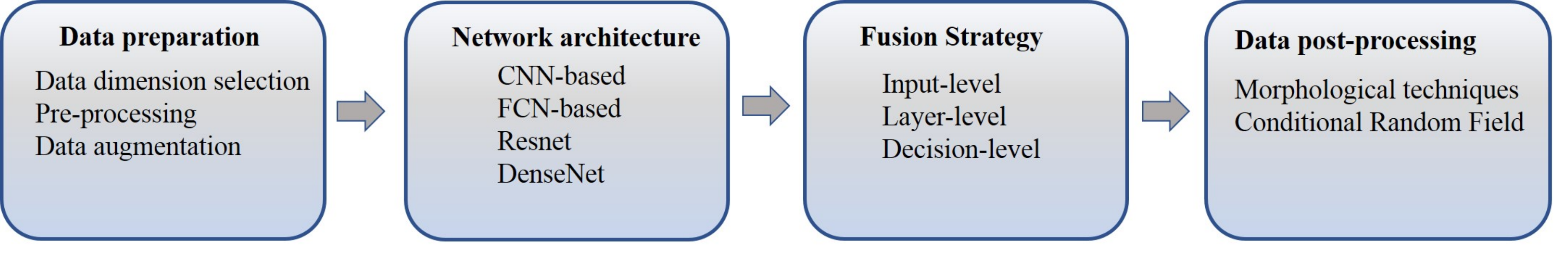} 
\caption{The pipeline of multi-modal medical image segmentation based on deep learning.}
\label{4}
\end{figure}

We describe a pipeline of multi-modal medical image segmentation based on deep learning, shown in Figure \ref{4}. The pipeline consists of four parts: data preparation, network architecture, fusion strategy and data post-processing. In the data preparation stage, the data dimension is firstly chosen, and the pre-processing is used to reduce the variation between images, and data augmentation strategy can also be used to increase the training data to avoid the over-fitting problem. In the network architecture and fusion strategy stages, the basic network and detailed multi-modal images fusion strategies are presented to train the segmentation network. In the data post-processing stage, some post-pressing techniques such as morphological techniques and conditional random field are implanted to refine the final segmentation result. In the task of multi-modal medical image segmentation, fusing multiple modalities is the key problem of the task. According to the level in the network architecture where the fusion is performed, the fusion strategies can be categorized into three groups: input-level fusion, layer-level fusion, and decision-level fusion, the details refers to Section 4.

\begin{table}[]
\begin{center}
\resizebox{\textwidth}{!}{%
\begin{tabular}{ccccccc}
\toprule
Dataset & Train & Validation & Test & Segmentation Task & Modality & Image Size \\ \midrule
Brats2012 & 35 & N/A & 15 & Brain tumor & T1, T1C, T2, Flair & 160$\times$216$\times$176 176$\times$176$\times$216 \\
Brats2013 & 35 & N/A & 25 & Brain tumor & T1, T1C, T2, Flair & 160$\times$216$\times$176 176$\times$176$\times$216 \\
Brats2014 & 200 & N/A & 38 & Brain tumor & T1, T1C, T2, Flair & 160$\times$216$\times$176 176$\times$176$\times$216 \\
Brats2015 & 200 & N/A & 53 & Brain tumor & T1, T1C, T2, Flair & 240$\times$240$\times$155 \\
Brats2016 & 200 & N/A & 191 & Brain tumor & T1, T1C, T2, Flair & 240$\times$240$\times$155 \\
Brats2017 & 285 & 46 & 146 & Brain tumor & T1, T1C, T2, Flair & 240$\times$240$\times$155 \\
Brats2018 & 285 & 66 & 191 & Brain tumor & T1, T1C, T2, Flair & 240$\times$240$\times$155 \\
ISLES2015 & 28 & N/A & 36 & Ischemic stroke lesion & T1, T2, TSE, Flair, DWI, TFE/TSE & 230$\times$230$\times$154 \\
 & 30 & N/A & 20 &  & T1c, T2, DWI, CBF, CBV, TTP, Tmax & N/A \\
MRBrainS13 & 5 & N/A & 15 & Brain Tissue & T1, T1\_1 mm, T1\_IR, Flair & 256$\times$256$\times$192 240$\times$240$\times$48 \\
NeoBrainS12 & 20 & N/A & 5 & Brain Tissue & T1, T2 & 384$\times$384$\times$50 512$\times$512$\times$110 512$\times$512$\times$50 \\
iSeg-2017 & 10 & N/A & 13 & Brain Tissue & T1, T2 & N/A \\
CHAOS & 20 & N/A & 20 & Abdominal Organs & CT, T1-DUAL, T2-SPIR & N/A \\
IVD & 16 & N/A & 8 & Intervertebral Disc & In-phase, Opposed-phase, Fat, Water & N/A \\ \bottomrule
\end{tabular}%
}
\end{center}
\caption{Summary of the multi-modal medical image segmentation datasets.}
\label{Table.2}
\end{table}

\begin{table}
\begin{center}
\resizebox{\textwidth}{!}{
\begin{tabular}{cc}
\toprule 
\textbf{Evaluation metric}& \textbf{Mathematical description}\\ 
\midrule
Dice score(DSC)& $DSC= \frac {2TP}{2TP+FP+FN}$\\
Sensitivity& $Sensitivity=\frac{TP}{TP+FN}$\\
Specificity& $Specificity =\frac{TN}{TN+FP}$\\
    Hausdorff distance(HD) & $HD =\max\{sup_{r_\in{\partial R}}d_m(s,r),sup_{s_\in\partial S}d_m(s,r)\}$\\
Absolute relative volume difference(ARVD)& $ARVD(X,Y)=\left|{100\times(\frac {\left|X\right|} {\left|Y\right|}-1)}\right| $\\
Average boundary distance (ABD)& $ABD(X_s,Y_s)=\frac{1}{N_{X_s}+N_{Y_s}}\Big(\sum_{x\in X_s} \min_{y\in Y_s}d(x,y)+\sum_{y\in Y_s} min_{x\in X_s}d(y,x)\Big)$\\
\bottomrule
\end{tabular}
}
\end{center}
\caption{Summary of the evaluation metrics commonly used for these datasets. with respect to the number of false positive ($FP$), true positive ($TP$), false negative ($FN$) and true negative ($TN$), $\partial S$ and $\partial R$ are the sets of lesion border pixels/voxels for the predicted and the truth segmentations, and $d_m(v,v)$ is the minimum of the Euclidean distances between a voxel $v$ and voxels in a set $v$. $\left|X\right|$ is the number of voxels in the reference segmentation and $\left|Y\right|$ is the number of voxels in the algorithm segmentation, $X_s$ and $Y_s$ are the sets of surface points of the reference and algorithm segmentations respectively. The operator $d$ is the Euclidean distance operator. }
\label{Table.3}
\end{table}

\section{Data processing}
This section will describe the data processing including data dimension selection, image pre-processing, data augmentation and post-processing techniques. This step is important in deep learning-based segmentation network.

\subsection{Data dimension}
Medical image segmentation usually deals with 3D images. Some models directly use the 3D images to train models \cite{isensee2017brain}\cite{isensee2018no}\cite{qin2018autofocus}\cite{dolz2018hyperdense}, while some models process the 3D image slice by slice \cite{pereira2016brain}\cite{cui2018automatic}\cite{dolz2018ivd}\cite{nie2016fully}\cite{wang2017automatic}. The 3D approach takes the 3D image as input and applies the 3D convolution kernel to exploit the spatial contextual information of the image. The main drawback is its expensive computational cost. Compared to utilizing the whole volume image to train the model, some 3D small patches can be used to reduce the computational cost. For instance, kamnitsas et al. \cite{kamnitsas2017efficient} extracts 10k random 3D patches at regular intervals for training to segment the brain lesion. The 2D approach takes the image slice or patch extracted from the 3D image as input and applies the 2D convolutional kernel, the 2D approach can efficiently reduce the computational cost, while it ignores the spatial information of the image in z direction. For example, Zhao, et al. \cite{zhao2018deep} trained firstly FCNNs using image patches and then CRFs as Recurrent Neural Networks using image slices with the parameters of FCNNs fixed, finally they fine-tuned the FCNNs and the CRF-RNN using image slices. To exploit the feature information of the 2D image and 3D image, Mlynarski, et al. \cite{mlynarski20193d} described a CNN-based model for brain tumor segmentation, it first extracts the 2D features of the image from axial, coronal and sagittal views and then takes them as the additional input of the 3D CNN-based model. The method can learn rich feature information in three dimensions, which achieve good performance with median Dice scores of 0.918 (whole tumor), 0.883 (tumor core) and 0.854 (enhancing core).

\subsection{Pre-processing}
Pre-processing plays an important role in subsequent segmentation task, especially for the multi-modal medical image segmentation because there are variant intensity, contrast and noise in the images. Therefore, to make the images appear more similar and make the network training smooth and quantifiable, some pre-processing techniques are applied before feeding to the segmentation network. The typical pre-processing techniques consist of image registration, bias field correction and intensity normalization. For BraTS dataset, the image registration has already done before provided to the public. \cite{pereira2016brain}\cite{cui2018automatic}\cite{kamnitsas2017efficient}\cite{zhao2018deep}\cite{kamnitsas2017ensembles} used the N4ITK method to correct the distortion of MRI data. \cite{havaei2017brain}\cite{pereira2016brain}\cite{isensee2017brain}\cite{cui2018automatic}\cite{wang2017automatic} \cite{kamnitsas2017efficient} proposed to normalize each modality of each patient independently by subtracting the mean and dividing by the standard deviation of the brain region.

\subsection{Data augmentation}
Most of the time, a large number of labels for training is not available for several reasons. Labelling the dataset requires an expert in this field which is expensive and time-consuming. When training large neural networks from limited training data, the over-fitting problem needs to be considered. \cite{perez2017effectiveness} Data augmentation is a way to reduce over-fitting and increase the amount of training data. It creates new images by transforming (rotated, translated, scaled, flipped, distorted and adding some noise such as Gaussian noise) the ones in training dataset. Both the original image and created images are fed into the neural network. For example, Isensee, et al. \cite{isensee2017brain} proposed to address over-fitting by utilizing a large variety of data augmentation techniques like random rotations, random scaling, random elastic deformations, gamma correction augmentation and mirroring on the fly during training.

\subsection{Post-processing}
\cite{hua2018multimodal} Post-processing is applied to refine the final result in segmentation network. The isolated segmentation labels with small size are prone to artefacts and the largest volume are usually kept in the final segmentation. In this case, morphological techniques are preferred to remove incorrect small fragments and keep the largest volume. And some post-processing techniques can be designed according to the structure of detected region. For example, considering LGG patients may don't have enhancing tumor, Isensee, et al. \cite{isensee2018no} proposed to replace all enhancing tumor voxels with necrosis if the number of predicted enhancing tumor is less than a threshold. Because if there is a false positive voxel in predicted segmentation where no enhancing tumor presents in the ground truth will result in a Dice score of 0. Another case in \cite{kamnitsas2017efficient}, a 3D fully connected Condition Random Field (CRF) is applied for post-processing to effectively remove false positives to refine the segmentation result. 

\section{Multi-modal segmentation networks}
\begin{figure}[htbp]
\centering
\includegraphics[width=12cm, height=3cm]{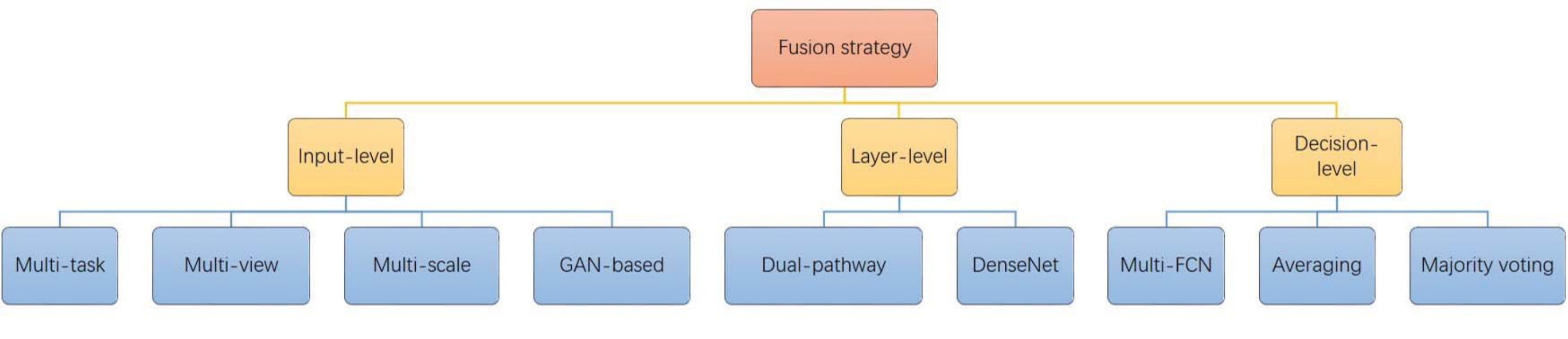} 
\caption{The generic categorization of the fusion strategy.}
\label{5}
\end{figure}

Over the years, various semi-automated and automated techniques have been proposed for multi-modal medical image segmentation using deep learning-based methods, such as CNN \cite{lecun1998gradient} and FCN \cite{long2015fully} especially U-Net \cite{ronneberger2015u}. According to the multi-modal fusion strategies, we category the network architectures into input-level fusion network, layer-level fusion network and decision-level fusion network, for each fusion strategy we conclude some common used menthods, shown in Figure \ref{5}.

\subsection{Input-level fusion network}
\begin{figure}[htbp]
\centering
\includegraphics[width=12cm, height=1.8cm]{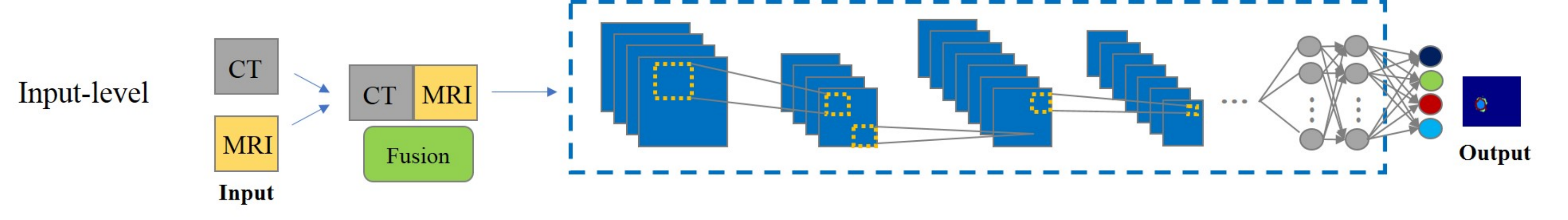} 
\caption{The generic network architecture of the input-level fusion.}
\label{6}
\end{figure}

In the input-level fusion strategy, multi-modality images are fused channel by channel as the multi-channel inputs to learn a fused feature representation, and then to train the segmentation network. Most of the existing multi-modal medical image segmentation networks adopt the input-level fusion strategy, which directly integrates the multi-modal images in the original input space \cite{pereira2016brain}\cite{isensee2017brain}\cite{isensee2018no}\cite{cui2018automatic}\cite{wang2017automatic}\cite{kamnitsas2017efficient}\cite{zhao2018deep}\cite{myronenko20183d}\cite{clerigues2018sunet}. Figure \ref{6} describes the generic network architecture of the input-level fusion segmentation network. We take CT and MRI as two input modalities, convolutional neural network as the segmentation network and the brain tumor segmentation as the segmentation task. By using the input-level fusion strategy, the rich feature information from different modalities can be fully exploited in all layers, from the first layer to the last one. This kind of fusion uses usually four techniques, multi-task segmentation, multi-view segmentation, multi-scale segmentation and GAN-based segmentation.

To name a few, Wang, et al. \cite{wang2017automatic} proposes a multi-modal segmentation network using BraTS dataset to segment the brain tumor into three subregions including the whole tumor, tumor core and enhancing tumor core. It uses multi-task and multi-view techniques. In order to obtain a united feature set, it directly integrates the four modalities (T1, T1c, T2 and Flair of MRI) as the multi-channel inputs in the input space. Then it separates the complex multi-class segmentation task into several simpler segmentation tasks according to the hierarchical structure of the brain tumor. The whole tumor is firstly segmented and then the bounding box including the whole tumor is used for the tumor core segmentation. Based on the obtained bounding box of the tumor core, the enhancing tumor core is finally segmented. Furthermore, to take advantage of 3D contextual information, for each individual task, they fused the segmentation results from three different orthogonal views (axial, coronal and sagittal) by averaging the softmax outputs of the individual task. Experiments with the testing set of BraTS 2017 data show that the proposed method achieves an average Dice scores of 0.7831, 0.8739, and 0.7748 for enhancing tumor core, whole tumor and tumor core, respectively, which won the second place on BraTS 2017 challenge. The multi-task segmentation separates the complex task of multiple class segmentation into several simpler segmentation tasks and takes advantage of the hierarchical structure of tumour subregions to improve segmentation accuracy.

Zhou et al. \cite{zhou2018one} also proposes a multi-task segmentation network on BraTS dataset, it fuses the multi-modal MR images channel by channel in the input space to learn a fused feature representation. Compared to segmentation of \cite{wang2017automatic} which suffers from network complexity and ignores the correlation between the three sequential segmentation tasks, it decomposes brain tumor segmentation into three different but related tasks. Each task has an independent convolutional layer, one classification layer, one loss layer and different input data. Based on curriculum learning \cite{bengio2009curriculum}, which means gradually increasing the difficulty of training tasks, they applied an effective strategy to improve the convergence quality of the model by training the first task only until the loss curve tends to flatten, then the first data and the second data are concatenated along the batch dimension as the input for the second task. The operation of the third task is like the second one. In this way, not only the model parameters but also the training data are transferred from an easier task to a more difficult task. The proposed approach ranks first on the BRATS 2015 test set and achieves top performance on the BRATS 2017 dataset.

It's likely to require different receptive field when segmenting different regions in an image. For example, large regions may need a large receptive field at the expense of fine details, while small regions may require high resolution local information. Qin et al. \cite{qin2018autofocus} proposed the autofocus convolutional layer to enhance the abilities of neural networks by using multi-scale processing. After integrating the multi-modal images in the input space, they applied an autofocus convolutional layer by using multiple convolutional layers with different dilation rates to change the size of the receptive field. Autofocus convolutional layer can indicate the importance of each scale when processing different locations of an image. Also, they used an attention mechanism to choose the optimal scale. The proposed autofocus layer can be easily integrated into existing networks to improve a model's performance. The proposed method gained promising performance on the challenging tasks of multi-organ segmentation in pelvic CT and brain tumor segmentation in MRI. 

Motivated by the success of Generative Adversarial Network (GAN)\cite{goodfellow2014generative}, which models a mini-max game between the generator and the discriminator, some methods propose to apply the discriminator as the extra constraint to improve the segmentation performance \cite{yang2018automatic}\cite{huo2018splenomegaly}. In \cite{yang2018automatic}, by fusing the multi-modal images as multi-channel inputs, they trained two separate networks: a residual U-net as the generative network and a discriminator network, the segmentation network will generate a segmentation, while the discriminator network will distinguish between the generated segmentations and ground truth masks. The discriminator is a shallow network containing three 3D convolution blocks, each followed by a max-pooling layer. In order to obtain a robust segmentation, they introduced extra constraints via contours to the model. Hausdorff distance between ground truth contours and prediction contours is used as a measure of dissimilarity. The proposed method was evaluated on the BraTS 2018 dataset and achieved competitive results, demonstrating that raw segmentation results can be improved by incorporating extra constraints in contours and adversarial training. Huo et al. \cite{huo2018splenomegaly} employed  the  PatchGAN \cite{isola2017image}  as  an additional discriminator to supervise the training procedure of the network. The method based on GAN can obtain a robust segmentation due to the extra constrain of discriminator, but it costs more memory to train the extra discriminator.

The input-level fusion strategy can maximumly keep the original image information and learn the intrinsic image feature. Using sequential segmentation networks allows to take different strategies, such as multi-task, multi-view, multi-scale and GAN-based segmentation network, to fully exploit the feature representation from multi-modal images.

\subsection{Layer-level fusion}
\begin{figure}[htbp]
\centering
\includegraphics[width=12.5cm, height=3.5cm]{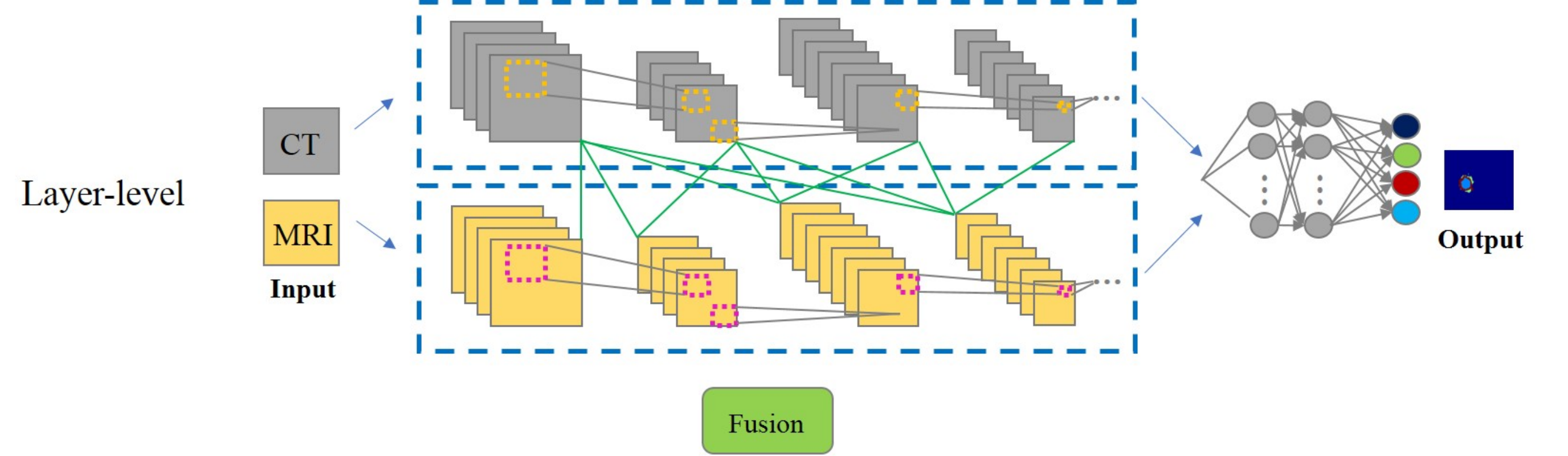} 
\caption{The generic network architecture of the layer-level fusion.}
\label{7}
\end{figure}

In the layer-level fusion strategy, single or two modal images are used as the single input to train individual segmentation network, and then these learned individual feature representations will be fused in the layers of the network, finally the fused result will be fed to the decision layer to obtain the final segmentation result. The layer-level fusion network can effectively integrate and fully leverage multi-modal images \cite{dolz2018hyperdense}\cite{dolz2018ivd}\cite{chen2018mri}\cite{rokach2010ensemble}. Figure \ref{7} describes the generic network
architecture of layer-level fusion segmentation work.

To name a few, we also take the brain tumor segmentation in multi-sequence of MRI to illustrate this kind of fusion. It is well known that T1 weighed MRI and T1c are suitable to segment the tumor core without the peritumoral edema, while T2 and Flair are suitable to segment the peritumoral edema. Chen et al. \cite{chen2018mri} proposes a dual-pathway multi-modal brain tumor segmentation network. The first pathway uses the T2 and Flair to extract the relative feature to segment the whole tumor from the background, and the second pathway uses the T1 and T1c to train the same segmentation network to learn other relative feature representation, and then the features from the both pathways are fused and finally fed into a four-class softmax classifier to segments the background, ED, ET and NCR/NET. The dual-pathway segmentation network can exploit the effective feature information of different modalities and achieve an accurate segmentation result.

Dolz, et al. \cite{dolz2018hyperdense} proposes a 3D fully convolutional neural network based on DenseNets that extends the definition of dense connectivity to multi-modal segmentation. Each imaging modality has a path and dense connections exist both in the layers within the same path and in the different paths. Therefore, the proposed network can learn more complex feature representations between the modalities. The extensive experiment results on two different and highly competitive multi-modal brain tissue segmentation challenges: iSEG 2017 \cite{wang2019benchmark} and MRBrainS 2013 \cite{mendrik2015mrbrains}, show that the proposed method yielded significant improvements over many other state-of-the-art segmentation networks, ranking at the top on both benchmarks.

Inspired by \cite{dolz2018hyperdense}, Dolz, et al.  \cite{dolz2018ivd} proposes an architecture for IVD (Intervertebral Disc) localization and segmentation in multi-modal MRI. Each MRI modality is processed in a corresponding single path to better exploit its feature representation. The network is densely-connected both within each path and across different paths, granting then the freedom of the model to learn where and how the different modalities should be processed and combined. It also improves the standard U-Net modules by extending inception modules using two convolutional blocks with dilated convolutions of a different scale to help handle multi-scale context information.

To summarize, in the layer-level fusion segmentation network, DenseNets are the commonly used networks which bring the three following benefits. First, direct connections between all layers help to improve the flow of information and gradients through the entire network, alleviating the problem of vanishing gradient. Second, short paths to all the feature maps in the architecture introduce implicit deep supervision. Third, dense connections have a regularizing effect, which reduces the risk of over-fitting on tasks with smaller training sets. Therefore, DenseNets allow to improved effectiveness and efficiency in the layer-level fusion segmentation network. In the layer-level fusion segmentation network, the connection among the different layers can capture complex relationships between modalities, which fully exploit the feature representation of multi-modal images.

\subsection{Decision-level fusion}
\begin{figure}[htbp]
\centering
\includegraphics[width=12.5cm, height=2.5cm]{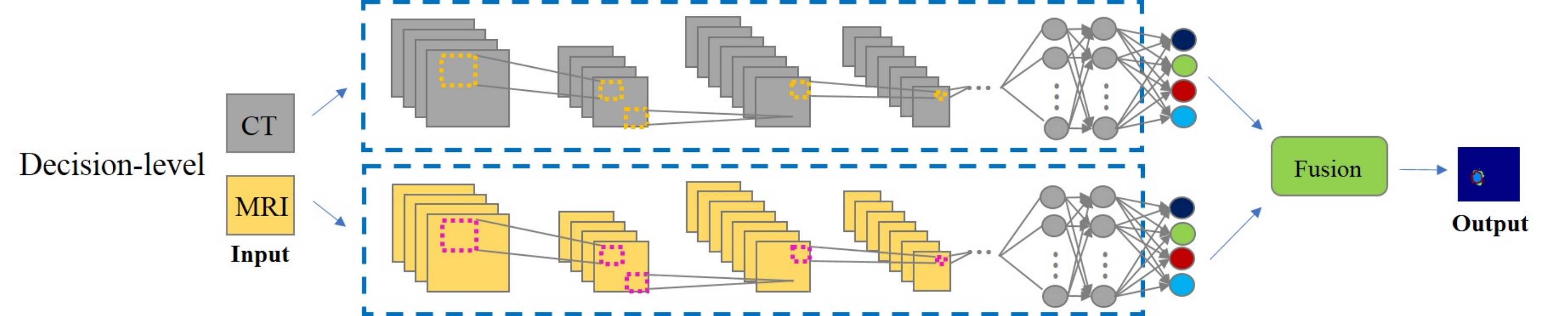} 
\caption{The generic network architecture of the decision-level fusion.}
\label{8}
\end{figure}

In decision-level fusion segmentation network, like the layer-level fusion, each modality image is used as the single input of single segmentation network. The single network can better exploit the unique information of the corresponding modality. The outputs of the individual networks will then be integrated to get the final segmentation result. The decision-level fusion segmentation network is designed to independently learn the complementary information from different modalities, since multi-modal images have little direct complementary information in their original image spaces due to different image acquisition techniques. Figure \ref{8} describes the generic network architecture of layer-level fusion segmentation work.

For example, to effectively employ multi-modalities from T1, T2 and fractional anisotropy (FA) modality, Nie, et al. \cite{nie2016fully} proposes a new multi-FCNs network architecture for the infant brain tissue segmentation (white matter (WM), gray matter (GM), and cerebrospinal fluid (CSF)). Instead of simply combining three modality data from the input space, they trained one network for each modality and then fused multiple modality features from high-layer of each network. Results showed that the proposed model significantly outperformed previous methods in terms of accuracy.

For the decision-level fusion, many fusion strategies have been proposed \cite{rokach2010ensemble}. The most of them are based on averaging and majority voting. For averaging strategy, Kamnitsas, et al. \cite{kamnitsas2017ensembles} trains three networks separately and then averaged the confidence of the individual networks. The final segmentation is obtained by assigning each voxel with the highest confidence. For majority voting strategy, the final label of a voxel depends on the majority of the labels of the individual networks.

The statistical properties of the different modalities are different, which make it difficult for a single model to directly find correlations across modalities. Therefore, in decision-level fusion segmentation network, the multiple segmentation networks can be trained to fully exploit the multi-modal features. Aygün, et al. \cite{aygun2018multi} investigates different fusion methods on the brain tumor segmentation problem in terms of memory and performance. In terms of memory usage, the decision-level fusion strategies require more memory since the model fuses the features later and more parameters are needed for layers to perform convolution and other operations. However, the later fusion can achieve better performance, because each modality is employed as input of one network which can learn complex and complementary feature information compared to input-level fusion network.

\section{Common problems}
\subsection{Over-fitting}
One limitation in medical image segmentation is data scarcity, usually leading to the over-fitting which refers to a model that has a good performance on the training dataset but does not perform well on new data. Most of the time, a large number of labels for training is not available for medical image analysis, because labelling the dataset requires experts in this field, and it is time-consuming and sometimes prone to error. When training complex neural networks with limited training data, special care must be taken to prevent the over-fitting. The complexity of a neural network model is defined by both its structure and the parameters. Therefore, we can reduce the complexity of the network architecture by reducing the layers or parameters or focus on methods that artificially increase the number of training data instead of changing the network architecture \cite{lecun2015deep}\cite{srivastava2014dropout}. The latter is commonly used to produce new synthetic images by performing data transformations and the corresponding ground truth that include operations of scaling, rotation, translation, brightness variation, elastic deformations, horizontal flipping and mirroring (for details refer to Section 3 data augmentation).

\subsection{Class imbalance}
One of the major challenges in medical image analysis is to deal with imbalanced data. In medical imaging field, the problem is even more glaring.  For example, for a segmentation of brain tumor or that of white matter lesion, the normal brain region is larger than the abnormal region. Training with the class imbalanced data can cause an instable segmentation network, which is biased towards the class with a large region. Table \ref{Table.4} illustrate the distribution of the classes in the training data of BraTS 2017, the number of positives (NEC/NET, ED and ET) and negatives (Background) are highly imbalanced and the background is overwhelmingly dominant. As result, the choice of the loss functions is crucial in segmentation networks, especially when dealing with highly unbalanced problems. We present several types of loss function which are widely used individually or combined in medical image segmentation networks. From the data-level, the problem can be addressed by resampling the data space. There are three main approaches: under-sampling the negative class \cite{jang2014medical}or up-sampling the negative class \cite{douzas2018effective}  and SMOTE (Synthetic Minority Over-sampling Technique) \cite{chawla2002smote} generating synthetic samples along the line segment that joins minority class samples. These approaches are simple to follow but they may remove some important data or add redundant data to the training set. 

\begin{table}
\centering
\label{}
\begin{center}
\begin{tabular}{c c c c c}
\hline
Region& Background&NET/NCR&Edema&Enhancing tumor\\ 
\hline
Percentage  &99.12&0.28 &0.40&0.20\\
\hline
\end{tabular}
\end{center}
\caption{The distribution of classes on BraTS 2017 training set, NET: Non Enhancing Tumor, NCR: Necrotic.}
\label{Table.4}
\end{table}

The patch sampling-based method can also mitigate the imbalanced data problem. For example, Kamnitsas, et al. \cite{kamnitsas2017efficient} proposes the balanced strategy to alleviate class imbalance problem. They extract the training patches with 50\% probability being cantered either on the lesion or healthy voxels. Clèrigues, et al. \cite{clerigues2018sunet} uses the lesion centered strategy, in which all training patches are extracted from the region centered on a lesion voxel. Additionally, a random offset is added to a sampling point to avoid location bias, where a lesion voxel is always expected at the patch center, contributing then to some data augmentations. 

As for the algorithm-level, Havaei, et al. \cite{havaei2017brain} proposes a two-phase training procedure.  It first constructs a patch dataset such that all labels are equiprobable by taking into account the diversity in all classes, and then retrains only the output layer to calibrate the output probabilities correctly. In this way, the class imbalance problem is overcome. Another approach consists of using a multi-task segmentation \cite{wang2017automatic}\cite{zhou2018one}\cite{chen2018mri}\cite{shen2017multi} that decomposes the complex multi-class segmentation task into several simple tasks, since each training task segments only one region, where the label distribution will be less unbalanced than segmentation multiple classes at one time.  Some approaches address class imbalance problem through an ensemble learning by combining same or different classifiers to improve their generalization ability \cite{sun2007cost}. Otherwise, the loss functions can also alleviate this problem by modifying the distributions of the training data. We present them as below. 

\noindent {\textbf {Cross-Entropy (CE) loss:}} Cross-entropy loss function is the most commonly used for the task of image segmentation. It is calculated by the equation (1). Because the cross-entropy loss evaluates individually the class predictions for each pixel vector and then averages all pixels, this can lead some error if an unbalanced class representation exists in the image. Long, et al. \cite{long2015fully} proposes to weight or sample the loss function for each output channel in order to alleviate the class imbalance problem. 
\begin{equation}
    \ Loss_{CE}=-\sum_{i\in N}\sum_{l\in L}y_i^{(l)}log \hat{y}_i^{(l)}
\end{equation}
Where $N$ is the set of all examples and $L$ the set of all labels, $y_i^{(l)}$ is the one-hot encoding (0 or 1) for example  and label l, $\hat{y}_i^{(l)}$ is the predicted probability for the same example/label pair.

\noindent {\textbf{Weighted Cross Entropy (WCE):}} Since the background regions dominate the training set, it is reasonable to incorporate the weights of multiple classes into the cross-entropy as defined as follows \cite{lecun2015deep}:
\begin{equation}
    \ Loss_{WCE}=-\sum_{i\in N}\sum_{l\in L}w_i y_i^{(l)}log \hat{y}_i^{(l)}
\end{equation}
Where $w_i$ represents the weight assigned to the $ith$ label. 

\noindent {\textbf{Dice loss (DL):}} Dice loss is a popular loss function for medical image segmentation which is a measure of overlap between the predicted sample and real sample. This measure ranges from 0 to 1 where a Dice score of 1 denotes the complete overlap as defined as follows \cite{milletari2016v}:
\begin{equation}
    \ Loss_{DL}=1-2\frac{\sum_{l\in L}\sum_{i\in N}y_i^{(l)}\hat{y}_i^{(l)}+\epsilon} {\sum_{l\in L}\sum_{i\in N}(y_i^{(l)}+\hat{y}_i^{(l)})+\epsilon}
\end{equation}
Where $\epsilon$ is a small constant to avoid dividing by 0.

\noindent {\textbf{Generalized Dice (GDL):}} Sudre, et al. \cite{sudre2017generalised} proposed to use the class re-balancing properties of the Generalized Dice overlap, defined in (4), as a robust and accurate deep-learning loss function for unbalanced tasks. The authors investigate the behavior of Dice loss, cross-entropy loss and generalized dice loss functions in the presence of different rates of label imbalance across 2D and 3D segmentation tasks. The results demonstrate that the GDL is more robust than the other loss functions.
\begin{equation}
    \ Loss_{GDL}=1-2\frac{\sum_{l\in L}w_i\sum_{i\in N} y_i^{(l)}\hat{y}_i^{(l)}+\epsilon} {\sum_{l\in L}w_i\sum_{i\in N} (y_i^{(l)}+\hat{y}_i^{(l)})+\epsilon}
\end{equation}

\noindent {\textbf{Focal loss (FL):}} Focal loss was originally introduced for the detection task. It encourages the model to down-weight easy examples and focuses training on hard negatives. Formally, the Focal loss is defined by introducing a modulating factor  to the cross-entropy loss and a parameter for class balancing \cite{lin2017focal}:  
\begin{equation}
 \ Loss_{{FL}_{(p_t)}}=-\alpha_t(1-p_t)^\gamma log(p_t)
\end{equation}

\begin{equation}
p_t=
\begin{cases}
p_t &\text{if y=1} \\
1-p_t &\text{otherwise}
\end{cases} 
\end{equation}
Where $y\in\{-1, +1\}$ is the ground-truth class, and $p_t\in[0, 1]$ is the estimated probability for the class with label $y=1$. The focusing parameter $\gamma$ smoothly adjusts the rate at which easy examples are down-weighted, setting $\gamma > 0$ can reduce the relative loss for well-classified examples, putting focus on hard and misclassified examples, the focal loss is equal to the original cross entropy loss when $\gamma =0$.

\vspace*{15pt}

\begin{table}[htp]\small
\begin{center}
\resizebox{\textwidth}{!}{%
\newcommand{\tabincell}[2]{\begin{tabular}{@{}#1@{}}#2\end{tabular}}
\begin{tabular}{ccccccc}
\toprule 
\textbf{Article}& \textbf{Pre-processing}& \textbf{Data }& \textbf{Network}& \textbf{Fusion level}& \textbf{Results (DSC)}& \textbf{Dataset}\\ 
\midrule 
\cite{kamnitsas2017efficient}*& \tabincell{c}{Normalization\\Bias Field Correction}& \tabincell{c}{3D\\Patch}& \tabincell{c}{CNN\\CRF}& Input& \tabincell{c}{whole/core/enhanced\\0.84/0.66/0.63}& \textbf{BraTS15}\\ 
\cite{pereira2016brain}& \tabincell{c}{Normalization\\Bias Field Correction}& \tabincell{c}{2D\\Patch}& CNN& Input& \tabincell{c}{whole/core/enhanced\\0.84/0.71/0.57}& BraTS13\\ 
\cite{isensee2017brain}*& \tabincell{c}{Normalization\\Data Augmentation}& \tabincell{c}{3D\\Patch}& \tabincell{c}{U-Net\\ResNet}& Input& \tabincell{c}{whole/core/enhanced\\0.85/0.74/0.64\\0.85/0.77/0.64}& \tabincell{c}{BraTS15\\BraTS17}\\ 
\cite{zhao2018deep}& \tabincell{c}{Normalization\\Bias Field Correction}& \tabincell{c}{3D\\Patch}& \tabincell{c}{FCN\\CRF\\RNN}& Input& \tabincell{c}{whole/core/enhanced\\0.86/0.73/0.62\\0.84/0.73/0.62\\4/3/2(rank)}& \tabincell{c}{BraTS13\\BraTS15\\BraTS16}\\ 
\cite{zhou2018one}& Normalization & 3D & \tabincell{c}{U-Net\\ResNet}& Input& \tabincell{c}{whole/ core/enhanced\\0.87/0.75/0.64}& BraTS15\\ 
\cite{wang2017automatic}& \tabincell{c}{Normalization\\Bias Field Correct} & \tabincell{c}{2D\\Slice}& \tabincell{c}{U-Net\\ResNet}& Input& \tabincell{c}{whole/core/enhanced\\0.87/0.77/0.78}& BraTS17\\ 
\cite{isensee2018no}& \tabincell{c}{Normalization\\Data Augmentation}& \tabincell{c}{3D\\Patch}& \tabincell{c}{U-Net\\ResNet}& Input& \tabincell{c}{whole/core/enhanced\\0.87/0.80/0.77}& BraTS18\\ 
\cite{cui2018automatic}& \tabincell{c}{Normalization\\Data Augmentation\\Bias Field Correction}& \tabincell{c}{2D\\Patch}& \tabincell{c}{CNN\\FCN}& Input& \tabincell{c}{whole/core/enhanced\\0.89/0.77/0.80}& BraTS15\\ 
\cite{pereira2016brain}& \tabincell{c}{Normalization\\Bias Field Correction}& 3D & \tabincell{c}{CNN}& Input& \tabincell{c}{whole/core/enhanced\\0.88/0.81/0.76}& \textbf{BraTS13}\\ 
\cite{chang2016fully}& \tabincell{c}{Normalization\\Data Augmentation}& 2D & \tabincell{c}{FCN}& Input& \tabincell{c}{whole/core/enhanced\\0.87/0.81/0.72}& \textbf{BraTS16}\\
\cite{kamnitsas2017ensembles}*& \tabincell{c}{Normalization\\Bias Field Correction}& 3D & \tabincell{c}{U-Net\\FCN\\DeepMedic}& Input& \tabincell{c}{whole/core/enhanced\\0.88/0.78/0.72}& \textbf{BraTS17}\\
\cite{myronenko20183d}*& \tabincell{c}{Normalization\\Data Augmentation}& 3D & \tabincell{c}{U-Net\\VAE}& Input& \tabincell{c}{whole/core/enhanced\\0.88/0.81/0.76}& \textbf{BraTS18}\\
\cite{georgiev2018automatic}& N/A & \tabincell{c}{2D\\Slice} & \tabincell{c}{CNN\\ResNet}&Input &0.9112 &\textbf{IVD}\\
\cite{clerigues2018sunet}*& Normalization & \tabincell{c}{3D\\Patch}& \tabincell{c}{U-Net\\ResNet}& Input& \tabincell{c}{$0.59 \pm 0.31$\\$0.84 \pm 0.10$}& \textbf{ISLES15}(SISS/SPES)\\ 
\cite{van2013automated}*& Normalization & \tabincell{c}{3D\\Patch} & \tabincell{c}{SVM}& Input& \tabincell{c}{CSF/WM/GM\\0.78 0.88 0.84}& \tabincell{c}{\textbf{MRBrainS13}}\\
\cite{dolz2018hyperdense}*& N/A & \tabincell{c}{3D\\Patch} & \tabincell{c}{CNN\\DenseNet}& Layer& \tabincell{c}{CSF/WM/GM\\0.95/0.91/0.90\\0.84/0.90/0.86}& \tabincell{c}{iSEG-2017\\MRBrainS13}\\
\cite{bui20173d}*& Normalization & \tabincell{c}{3D\\Patch} & \tabincell{c}{3D\\DenseNet}& Layer& \tabincell{c}{CSF/WM/GM\\0.96/0.91/0.91}& \tabincell{c}{\textbf{iSEG-2017}}\\
\cite{dolz2018ivd}& N/A & \tabincell{c}{2D\\Slice} & \tabincell{c}{U-Net\\DenseNet}&Layer &$0.9191 \pm 0.0179$& IVD\\ 
\cite{nie2016fully}& N/A & \tabincell{c}{2D\\Patch} & FCN & Decision& \tabincell{c}{CSF/WM/GM\\
0.85/0.88/0.87}& Private data\\

\bottomrule 
\end{tabular}}
\end{center}
\caption{Summary of the deep learning approaches for multi-modal medical image
segmentation, the bold presents the best performance in the challenge. The acronyms in results are: cerebrospinal fuild(CSF), grey matter(GM), white matter(WM), the symbol * indicates the method has available code.}
\label{Table.5}
\end{table}

\section{Discussion and conclusion}
In the above sections, we presented a large set of state-of-the-art multi-modal medical image segmentation networks based on deep learning. They are summarized in Table \ref{Table.5}. For BraTS challenge, these methods are concluded since 2013, because deep learning methods are applied since 2013. Publicly available multi-modal medical image datasets for segmentation task are rare, the most used dataset is the BraTS dataset having proposed since 2012. For their segmentation, the current best method is proposed in \cite{myronenko20183d}, they use the input-level fusion strategy to directly integrate the different modalities in the input space, they apply the encoder-decoder structure of CNN combined with an additional VAE (variational autoencoder) branch to the encoder part. The VAE branch can reconstruct the input image and exploit better the features of the encoder endpoint. It also provides an additional guidance and a regularization to the encoder part. The authors demonstrate that more sophisticated data augmentation techniques, data post-processing techniques, or deeper network will not further improve the network performance, which means the network architecture plays a crucial role in the segmentation network than other data processing operations.

For multi-modal medical image segmentation, the fusion strategy takes an important role in order to achieve an accurate segmentation result. Conventional image fusion strategy learns a direct mapping between source images and target images, the fusion strategy consists of two basic stages: activity level measurement and fusion rule \cite{liu2017multi}. Activity level measurement is implemented by designing local filters to extract high-frequency details, and the calculated clarity information of different source images are then compared using some designed rules to obtain a clarity image. To achieve better performance, these issues become more and more complicated, so it is difficult to manually propose an ideal fusion strategy which fully concerns the important issues. To this end, a deep learning-based network can directly encode the mapping. 
Deep learning-based methods outperform in three aspects. First, deep learning-based networks learn a complex and abstract hierarchical feature representation for image data to overcome the difficulty of manual feature design. Second, deep learning-based networks can present the complex relationships between different modalities by using the hierarchical network layer, such as the layer-level fusion strategy. Third, the image transform and fusion strategy in the conventional fusion strategy can be jointly generated by training a deep learning model, in this way some potential deep learning network architectures can be investigated for designing an effective image fusion strategy. Therefore, the deep learning-based method has a great potential to produce better fusion results than conventional methods.

Choosing an effective deep learning fusion strategy is still an important issue. In 2013-2018 BraTS Challenge, all the methods applied the input-level fusion to directly integrate the different MR images in the input space, which is simple and can remain the intrinsic image feature and allow the method to focus on the subsequent segmentation network architecture designs, such as multi-task, multi-view, multi-scale and GAN-based strategies. While the strategy just concatenates the modalities in the input space, but it does not exploit the relationships among the different modalities. For layer-level fusion, with the dense connection among the layers, the fusion strategy often takes the DenseNet as the basic network. The connection among the different layers can capture complex relationships between modalities, which can help the segmentation network learn more valuable information and achieve better performance than directly integrating different modalities in the input space. For decision-level fusion strategy, it can achieve better performance compared to the input-level fusion, because each modality is employed to train a single network to learn independent feature representation, while this requires much memory and computational time. Compared the last two fusion strategies, the layer-fusion strategy seems better, since the dense connection among the layers can exploit more complex and complementary information to enhance the network training, while the decision-level fusion only learns the independent feature representation in single modality. Since the results of the three fusion strategies are not obtained from the same data, their comparison in terms of performance is difficult. Methodologically, each strategy has its advantages and disadvantages.

Although we observed the advantages of these fusion strategies based on deep learning, based on the previous works, we can still observe that there are some locks to lift in multi-modal medical image segmentation based on deep learning. It is known that multi-modal fusion networks generally perform better than single-modal network for segmentation task. The problem is how to fuse different modalities to get the best compromise for a precise segmentation. Hence, how to design multi-modal networks to efficiently combine different modalities, how to exploit the latent relationship between different modalities, and how to integrate the multi-information into the segmentation network to improve the segmentation performance can be the topics of future works. 

Other problem concerns the data. First, since it is difficult to obtain a large number of medical image data, the limited training data can easily lead to over-adjustment. To deal with it, reducing the complexity of the network architecture or increasing the number of training data has been proved to alleviate the problem. Second, training with the imbalanced data can cause an instable segmentation network especially with small lesion or structure segmentation. Resampling the data space, using two-phase training procedure, careful path sampling and appropriate loss function are the proposed strategies to overcome the problem. Third, like common problems for deep learning, it's difficult to train a deep network with original limited data without data augmentation or other optimized techniques. Therefore, designing faster methods to perform convolutions and appropriate optimization methods can help to train an effective segmentation network. It is becoming a widespread practice in the computer vision community to release source codes to the public. We have indicated the available code in Table 5. This practice helps to expedite the research in the field. Another recommended practice is validating the model on different datasets, which can open the door to design a robust model that can be applied to datasets of similar applications.





\vfill
\pagebreak
\bibliographystyle{splncs04}
\bibliography{strings}

\end{document}